\documentstyle[12pt,epsfig]{article}
\author{Karina I. Mazzitello$^{a}$, Juli\'an Candia$^{b}$, and V\'{\i}ctor Dossetti$^{c}$\\{}\\
$^a${\small\it Consortium of the Americas for Interdisciplinary Science and
Department of}\\{\small\it Physics and Astronomy,
University of New Mexico, Albuquerque, NM 87131, USA}\\
{\small\it CONICET and Departamento de F\'{\i}sica, Facultad de Ingenier\'{\i}a,}\\  
{\small\it Universidad Nacional de Mar del Plata, Mar del Plata, Argentina}\\
{\small Email address: kmazzite@mdp.edu.ar}\\
$^b${\small\it Consortium Americas for Interdisciplinary Science and
Department of}\\{\small\it Physics and Astronomy,
University of New Mexico, Albuquerque, NM 87131, USA}\\
{\small\it Center for Complex Network Research and Department of Physics,}\\  
{\small\it University of Notre Dame, Notre Dame, IN 46556, USA}\\
{\small Email address: jcandia@nd.edu}\\
$^c${\small\it Consortium of the Americas for Interdisciplinary Science and
Department of}\\{\small\it Physics and Astronomy,
University of New Mexico, Albuquerque, NM 87131, USA}\\
{\small Email address: dossetti@unm.edu}}

\title{Effects of Mass Media and Cultural Drift\\ in a Model for Social Influence} 

\begin{document}
\maketitle

\begin{abstract}
In the context of an extension of Axelrod's model for social influence,  
we study the interplay and competition between the cultural drift, represented as random perturbations, 
and mass media, introduced by means of an external homogeneous field.
Unlike previous studies [J. C. Gonz\'alez-Avella {\it et al},  Phys. Rev. E {\bf 72}, 065102(R) (2005)],
the mass media coupling proposed here is capable of affecting 
the cultural traits of any individual in the society, including those who do not share any features with 
the external message. A noise-driven transition is found:
for large noise rates, both the ordered (culturally polarized) phase and 
the disordered (culturally fragmented) phase are observed, while, for lower noise rates, the ordered 
phase prevails. In the former case, the external field is found to induce cultural ordering, 
a behavior opposite to that reported in previous studies using a different prescription for the mass 
media interaction. We compare the predictions of this model    
to statistical data measuring the impact of a mass media vasectomy promotion campaign in Brazil.
\end{abstract}

\vspace{0.4 true cm}

The non-traditional application of statistical physics to many problems of interdisciplinary nature
has been growing steadily in recent years. Indeed, it has been recognized that the 
study of statistical and complex systems can provide valuable tools and insight into 
many emerging interdisciplinary fields of science \cite{wei91,wei00,oli99}.  
In this context, the mathematical modeling of social phenomena allowed to perform quantitative investigations 
on processes such as self-organization, opinion formation and spreading, cooperation, formation 
and evolution of social structures, etc (see e.g. \cite{szn00,zan02,kup02,ale02,gonz06,can06}). 
In particular, a model for social 
influence proposed by Axelrod \cite{axe97a,axe97b}, which aims at understanding the formation of cultural 
domains, has recently received much attention \cite{cas00,vil02,kle03a,kle03b,gon05,gon06,kup06} 
due to its remarkably rich dynamical behavior.  

In Axelrod's model, culture is defined by the set of cultural attributes (such as language, art, 
technical standards, and social norms \cite{axe97b}) subject to social influence. The cultural state of 
an individual is given by their set of specific traits, which are  
capable of changing due to interactions with their acquaintances. In the original proposal, the individuals are 
located at the nodes of a regular lattice, and the interactions are assumed to take place between lattice neighbors.  
Social influence is defined by a simple local dynamics, which is assumed to satisfy the following two properties: 
(a) social interaction is more likely taking place between individuals that share some or many of their 
cultural attributes; (b) the result of the interaction is that of increasing the cultural similarity 
between the individuals involved.   

By means of extensive numerical simulations, it was shown that the system undergoes a 
phase transition separating an ordered (culturally polarized) phase from a disordered (culturally fragmented) one, 
which was found to depend on the number of different cultural traits available \cite{cas00}.  
The critical behavior of the model was also studied in different complex network topologies, such as small-world and 
scale-free networks \cite{kle03b}. 
These investigations considered, however, zero-temperature dynamics that 
neglected the effect of fluctuations. 

Following Axelrod's original idea of incorporating random perturbations to describe the effect 
of {\it cultural drift} \cite{axe97a}, noise was 
later added to the dynamics of the system \cite{kle03a}. 
With the inclusion of this new ingredient, the disordered multicultural configurations were found to be metastable 
states that could be driven to ordered stable configurations. The decay of 
disordered metastable states depends on the competition between the noise rate, $r$, and the 
characteristic time for the relaxation of perturbations, $T$. Indeed, for $r\leq T^{-1}$, the perturbations drive  
the disordered system towards monocultural states, while, for $r\geq T^{-1}$, the noise 
rates are large enough to hinder the relaxation processes, thus keeping the disorder. Since $T$ scales with the 
system size, $N$, as $T\sim N{\rm ln}N$, the culturally fragmented states persist in the thermodynamic limit, 
irrespective of the noise rate \cite{kle03a}. 

More recently, an extension of the model was proposed, in 
which the role of {\it mass media} and other mass external agents 
was introduced by considering external \cite{gon05} and autonomous local or global 
fields \cite{gon06}, but neglecting random fluctuations.
The interaction between the fields and the individuals was chosen to resemble the 
coupling between an individual and their neighbors in the original Axelrod's model. According to the adopted 
prescription, the interaction probability was assumed to be null 
for individuals that do not share any cultural feature with the external message. In this way, 
intriguing, counterintuitive 
results were obtained: the influence of mass media was found to disorder the 
system, thus driving ordered, culturally polarized states towards 
disordered, culturally fragmented configurations \cite{gon05}. 

The aim of this work is to include the effect of cultural drift in an alternative mass media scenario.
Although still inspired in the original Axelrod's interaction, the mass media coupling proposed here is 
capable of affecting the cultural traits of any individual in the society, including those who do not share any features with 
the external message. 

For noise rates below a given transition value, which depends on the intensity of the mass media interactions, 
only the ordered phase is observed.
However, for higher levels of noise above the transition perturbation rate, both the ordered (culturally polarized) phase and 
the disordered (culturally fragmented) phase are found. 
In the latter case, we obtain an order-disorder phase diagram as a function of the field intensity and the number of 
traits per cultural attribute. According to this phase diagram, the role of the external field is that of inducing cultural ordering, 
a behavior opposite to that reported in Ref.~\cite{gon05} using a different prescription for the mass 
media interaction. In order to show the plausibility of the scenario considered here, 
we also compare the predictions of this model
to statistical data measuring the impact of a mass media vasectomy promotion campaign in Brazil \cite{kin96}.

The model is defined by considering individuals located at the sites of an $L\times L$ square lattice. The cultural 
state of the $i-$th individual is described by the integer vector $(\sigma_{i1},\sigma_{i2},...,\sigma_{iF})$, where 
$1\leq\sigma_{if}\leq q$. The dimension of the vector, $F$, defines the 
number of cultural attributes, while $q$ corresponds to the number of different cultural traits per attribute. 
Initially, the specific traits for each individual are assigned randomly with a uniform distribution. Similarly,  
the mass media cultural message is modeled by a constant integer vector 
$(\mu_1,\mu_2,...,\mu_F)$, which can be chosen as $(1,1,...,1)$ without loss of generality. 
The intensity of the mass media message relative to the local interactions between neighboring 
individuals is controlled by the parameter $M$ ($0\leq M\leq 1$). Moreover, the parameter $r$ ($0\leq r\leq 1$) 
is introduced to represent the noise rate \cite{kle03a}. 

The model dynamics is defined by iterating a sequence of rules, as follows: (1) an individual
is selected at random; (2) with probability $M$, he/she interacts with the mass media field;  
otherwise, he/she interacts with a randomly chosen nearest neighbor;
(3) with probability $r$, a random single-feature perturbation is performed.  

The interaction between the $i-$th and $j-$th individuals is governed by their cultural overlap,  
$C_{ij}=\sum_{f=1}^F\delta_{\sigma_{if},\sigma_{jf}}/F$, where $\delta_{kl}$ is the Kronecker delta. 
With probability $C_{ij}$, the result of the interaction is that of 
increasing their similarity: one chooses at random one of the attributes on which they differ 
(i.e., such that $\sigma_{if}\neq\sigma_{jf}$) and sets them equal by changing the trait of 
the individual selected in first place. Naturally, if $C_{ij}=1$, 
the cultural states of both individuals are already identical, and the interaction leaves them unchanged. 

The interaction between the $i-$th individual and the mass media field is governed by the overlap term 
$C_{iM}=(\sum_{f=1}^F\delta_{\sigma_{if},\mu_f}+1)/(F+1)$. Analogously to the precedent case, 
$C_{iM}$ is the probability that, as a result of the interaction, the individual changes one of the traits 
that differ from the message by setting it equal to the message's trait. 
Again, if $C_{iM}=1$, the cultural state of the individual is already identical to the mass media message, 
and the interaction leaves it unchanged. 
Notice that $C_{iM}>0$; thus, the mass media coupling used here is 
capable of affecting the cultural traits of any individual in the society, 
including those who do not share any features with the external message. 
As commented above, this differs from the mass media interaction proposed in Ref.~\cite{gon05}, 
which was given by $C'_{iM}=\sum_{f=1}^F\delta_{\sigma_{if},\mu_f}/F$.

\begin{figure}[t]
\begin{center}
\epsfxsize=4.2truein\epsfysize=3.1truein\epsffile{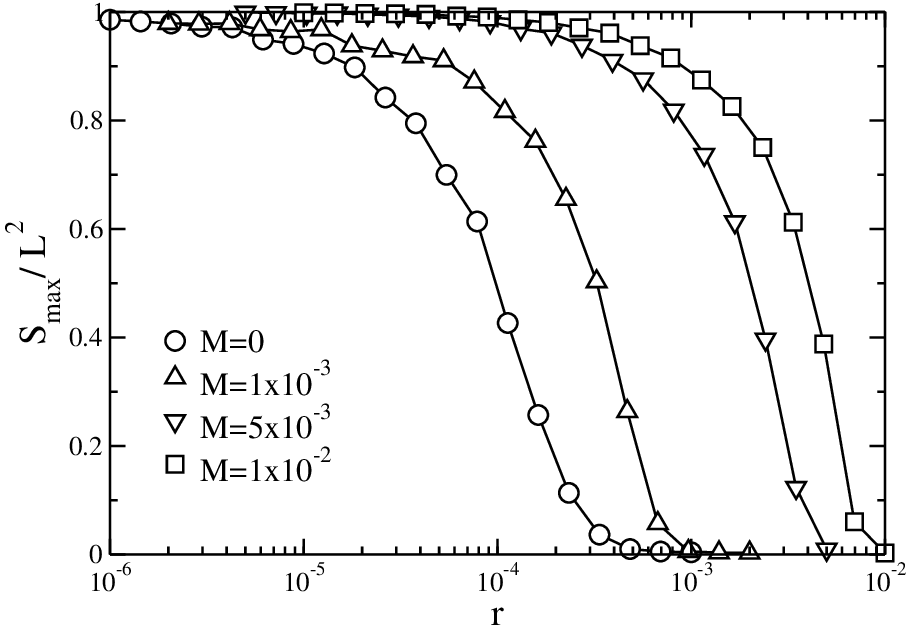}
\caption{Order parameter, $S_{max}/L^2$, as a function of the noise rate, $r$, for different 
values of the mass media intensity $M$, as indicated. 
The number of different cultural traits per attribute is $q=40$. 
The lines are guides to the eye.}
\label{fig1}
\end{center}
\end{figure}

As regards the perturbations introduced in step (3),
a single feature of a single individual is randomly chosen, and, with probability $r$, 
their corresponding trait is changed to a randomly selected value between 1 and $q$. 

In the absence of fluctuations, the system evolves towards absorbing states, i.e., frozen configurations that 
are not capable of further changes. For $r>0$, instead, the system evolves continuously, and, after a transient period, 
it attains a stationary state. In order to characterize the degree of order of these stationary states, 
we measure the (statistically-averaged) size of the largest homogeneous domain, $S_{max}$ \cite{cas00,kle03a}. 
The results obtained here correspond to systems of linear size $L=50$ and a fixed number of cultural attributes, $F=10$, 
typically averaged over 500 different (randomly generated) initial configurations.    

\begin{figure}[t]
\begin{center}
\epsfxsize=4.2truein\epsfysize=3.1truein\epsffile{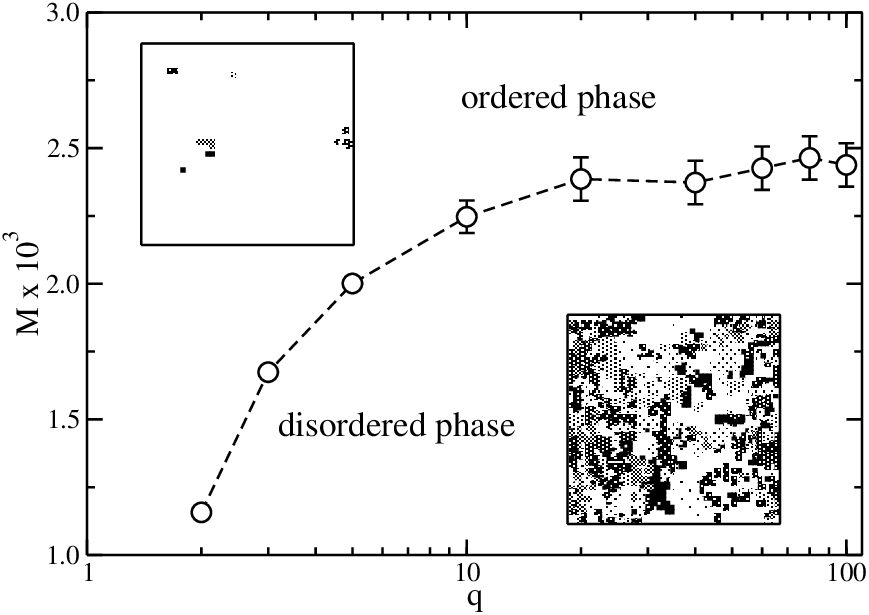}
\caption{Phase diagram showing the separation between the ordered (culturally polarized) phase 
and the disordered (culturally fragmented) phase, for the noise rate $r=10^{-3}$. The dashed line is 
a guide to the eye. Typical snapshot configurations of both phases are also shown, where the cultural 
state corresponding to the external message is indicated in white.}
\label{fig2}
\end{center}
\end{figure}

Figure 1 shows the order parameter, $S_{max}/L^2$, as a function of the noise rate, $r$, for different 
values of the mass media intensity. The number of different cultural traits per attribute is $q=40$. 
As anticipated, for small noise rates, the perturbations drive the decay of disordered 
metastable states, and thus the system presents only ordered states with $S_{max}/L^2\approx 1$.  
As the noise rate is gradually increased, the competition between characteristic times for perturbation 
and relaxation processes sets on, and, for large enough noise rates, the system becomes completely disordered. 
This behavior, which was already reported in the absence of mass media interactions \cite{kle03a}, is here 
also observed for $M>0$. As we consider plots for increasing values of $M$,
the transition between ordered and disordered states takes place for increasingly higher levels of noise. 
Indeed, this is an indication of the competition between 
noise rate and external field effects, thus showing that the external field induces order in the system.  

Figure 2 shows the order-disorder phase diagram as a function of the field intensity and the number of 
traits per cultural attribute, for the noise rate $r=10^{-3}$. The transition points correspond to $S_{max}/L^2=0.5$. 
For the $M=0$ case, noise-driven order-disorder transitions were found to be roughly independent 
of the number of traits per cultural attribute, as long as $q\geq 10$ \cite{kle03a}. Here, we observe a similar, essentially 
$q-$independent behavior for $M>0$ as well. Typical snapshot configurations of both regions are also shown in Figure 2, where 
the transition from the (small-$M$) multicultural regime to the (large-$M$) monocultural state is clearly observed. 
A majority of individuals sharing the same cultural state, identical to the external message, is found within the ordered phase. 
For smaller noise rates, $r\leq 10^{-4}$, the system is ordered even for $M=0$, and hence only the monocultural 
phase is observed. 

\begin{figure}[t]
\begin{center}
\epsfxsize=4.2truein\epsfysize=3.1truein\epsffile{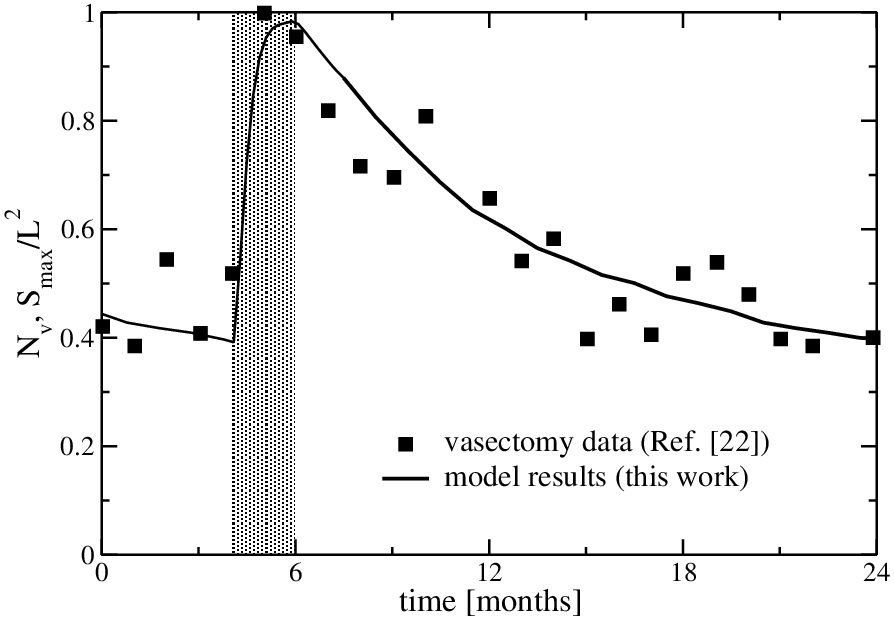}
\caption{Comparison of model predictions (solid line) to statistical data (symbols) measuring the 
impact of a mass media vasectomy promotion campaign in Brazil \cite{kin96}. The number of vasectomies per month, $N_v$,
has been normalized and compared to the model's order parameter, $S_{max}/L^2$, while the time scales have been 
matched by setting 1 month = 500 MCS. The shaded region indicates the 
time window in which the mass media campaign was performed.}
\label{fig3}
\end{center}
\end{figure}

In order to gain further insight into the interplay and competition between cultural drift and mass media effects, 
let us now consider the external message being periodically switched on and off. Starting with a random disordered 
configuration and assuming a noise level above the transition value for the $M=0$ case, 
we observe a periodical behavior: the system 
becomes ordered within the time window in which the field is applied, while it becomes increasingly disordered when 
the message is switched off. A cycle representing this behavior is shown by the solid line in Figure 3, which 
corresponds to $r=1.5\times 10^{-4}$, $M=10^{-2}$, and $q=40$. 

Moreover, we can compare this behavior to statistical data 
measuring the impact of a mass media vasectomy promotion campaign in Brazil \cite{kin96}. Symbols in Figure 3 correspond to 
the number of vasectomies performed monthly in a major clinic in S\~ao Paulo, spanning a time interval of 
2 years. The shaded region indicates the time window in which the mass media campaign was performed. 
The promotion campaign consisted of prime-time television and radio spots, the distribution of flyers, an 
electronic billboard, and public relations activities.  
In order to allow a comparison to model results, vasectomy data have been normalized by setting the 
maximal number of vasectomies measured equal to unity, while the relation between time scales has been chosen conveniently.    
In the model results, 
time is measured in Monte Carlo steps (MCS), where 1 MCS corresponds to $L^2$ iterations of the set of rules (1)-(3). 
For the comparison performed in Figure 3, we assumed that 1 month corresponds to 500 MCS. 
Although the model parameters and scale units were arbitrarily assigned, it is reassuring to observe that a good 
agreement between observations and model results can be achieved. Indeed, 
the steep growth in the number of vasectomies practiced during the promotion campaign, as well as the 
monotonic decrease afterwards, can be well accounted for by this model. 

In order to carry out a straightforward comparison between mass media effects in this model and 
the measured response within a social group, several simplifying assumptions were adopted. As commented 
above, the model parameters and scale units were conveniently assigned. Moreover, no distinction was 
attempted between opinions (as modeled, within Axelrod's representation, by sets of cultural 
attributes in the mathematical form of vectors) and actual choices (as measured e.g. by the vasectomy 
data shown in Figure 3). In related contexts, analogous simplifying assumptions were adopted in the 
statistical physics modeling of political phenomena (e.g. the distribution of votes in elections 
in Brazil and India \cite{ber01,ber02,gon04}, Italy and Germany \cite{car05}, four-party political 
scenarios \cite{szn05}, etc), marketing competition 
between two advertised products \cite{sch03,szn03}, and applications to finance \cite{szn02}.      

In summary, we have studied, in the context of an extension of Axelrod's model for social influence,   
the interplay and competition between cultural drift and mass media effects. 
The cultural drift is modeled by random perturbations, while mass media effects are introduced by means of an external field. 

A noise-driven order-disorder transition is found. In the large noise rate regime, both the ordered (culturally polarized) phase 
and the disordered (culturally fragmented) phase can be observed, whereas in the small noise rate regime, only the ordered 
phase is present.
In the former case, we have obtained the corresponding order-disorder phase diagram, showing
that the external field induces cultural ordering. This behavior is opposite to that reported in Ref.~\cite{gon05} 
using a different prescription for the mass media field, which neglected the interaction between the field and individuals 
that do not share any features with the external message. The mass media coupling proposed in this work, instead, is capable 
of affecting the cultural traits of any individual in the society.  

In order to show the plausibility of the scenario considered here, 
we have compared the predictions of this model
to statistical data measuring the impact of a mass media vasectomy promotion campaign in Brazil. 
A good agreement between model results and measured data can be achieved. The observed behavior is characterized by 
a steep growth during the promotion campaign, and a monotonic decrease afterwards.  
We can thus conclude that the extension of Axelrod's model proposed 
here contains the basic ingredients needed to explain the trend of actual observations. 

We hope that the present findings will contribute to the growing interdisciplinary efforts 
in the mathematical modeling of social dynamics phenomena, and stimulate further work.

\section*{Acknowledgments}
We acknowledge useful and stimulating discussions 
with M. G. Cosenza and V. M. Kenkre. This work was supported by the NSF under grant no. INT-0336343.
J. C. is also supported by the James S. McDonnell Foundation and the National 
Science Foundation ITR DMR-0426737 and CNS-0540348 within the DDDAS program.

\end{document}